\documentclass[%
 reprint,
 amsmath,amssymb,
 aps,
]{revtex4-2}

\usepackage{graphicx}% Include figure files
\usepackage{dcolumn}% Align table columns on decimal point
\usepackage{bm}% bold math
\usepackage[usenames]{color}

\begin{document}

\preprint{APS/123-QED}

\title{Modification of Damon-Eshbach magnetostatic mode spectra in ferromagnet/paramagnet bilayer}

\author{M. A. Kuznetsov}

\email{kuznetsovm@ipmras.ru}%Lines break automatically or can be forced with \\

\affiliation{%
Institute for Physics of Microstructures, Russian Academy of Sciences, Nizhny Novgorod 607680, Russian Federation
}

\date{\today}% It is always \today, today,

\begin{abstract}
Using the magnetostatic approximation, we calculate the spectra of bulk and surface spin waves in an in-plane magnetized ferromagnet/paramagnet bilayer. Due to the dipolar coupling between the layers, the paramagnet becomes polarized, which in turn modifies the spectrum of the Damon-Eshbach magnetostatic modes. We assume that the paramagnet is characterized by a magnetic susceptibility, $\chi \propto 1/(T-T_C)$, which reaches large values when the system temperature $T$ is close to the Curie temperature  $T_C$. We find the conditions under which surface spin waves become unidirectional, i.e., capable of carrying energy in only one direction, and determine the magnitude of their frequency nonreciprocity. We demonstrate the possibility of switching the unidirectional wave regime on and off by varying the external magnetic field or temperature, making the ferromagnet/paramagnet system an attractive platform for tunable magnonic logic devices.
\end{abstract}

\maketitle

\section{Introduction}
\label{sec:intro}

The development of magnonic devices, where information transmission and processing are carried out via spin waves (magnons), has attracted significant interest because such devices offer several advantages over their conventional electronic counterparts. These advantages include the absence of Joule heating, the tunability of spin-wave spectra via an external magnetic field and system geometry, and the micro- and submicrometer wavelengths at microwave frequencies, which facilitates device miniaturization \cite{kruglyak_magnonics_2010,Mahmoud_2020,flebus_2024_2024,han_magnonics_2024}. These features make magnonics a promising and rapidly advancing field of electronics \cite{chumak_advances_2022}. Typically, in-plane magnetized ferromagnetic (FM) or ferrimagnetic films serve as spin-wave buses, with their magnetostatic modes originally described by Damon and Eshbach \cite{damon_magnetostatic_1961}. In such films, both surface (MSSWs) and backward-volume (BVMSWs) magnetostatic waves can propagate. To date, numerous studies have demonstrated that the properties of spin waves (spectrum, group velocity, propagation length, etc.) can be modified by coupling the FM film to another material, such as a normal metal \cite{seshadri_surface_1970,De_Wames_1970,Mruczkiewicz_2014}, superconductor \cite{Golovchanskiy_2018,Kuznetsov_2022,Kharlan_2024,zhou_giant_2024}, heavy metal \cite{udvardi_chiral_2009,moon_spin-wave_2013,Di_2015,Di_2015_2}, another FM film \cite{Wolfram_1970,camley_magnetostatic_1982,Zhang_1987,Gallardo_2019,Song_2020,heins_nonreciprocal_2025}, or a paramagnet \cite{Kuznetsov_2022,Vashkovskii_2006}.

Among the diverse materials considered for modifying spin-wave properties, paramagnets (PMs) remain relatively underexplored, despite giving rise to unusual effects. For instance, it has been shown that dipolar coupling to a PM substrate breaks the chiral symmetry of the static magnetization distribution in the FM film \cite{Mikuszeit_2011,Fraerman_2020,kuznetsov_magnetostatic_2023} and takes the form of an effective Dzyaloshinskii-Moriya interaction \cite{Kuznetsov_2023}. A consequence of this effective interaction is a strong frequency nonreciprocity of spin waves propagating in ultrathin FM film when their amplitude is uniform across the film thickness \cite{Kuznetsov_2022}. In Ref. \cite{Vashkovskii_2006}, magnetostatic modes in an in-plane magnetized FM film were studied subject to a ``magnetic wall'' boundary condition, which requires the tangential component of the magnetic field strength to vanish. This condition is equivalent to the FM film being in dipolar contact with an ideal paramagnet possessing an infinitely large susceptibility $\chi$. It was shown that under such a boundary condition, MSSWs become unidirectional meaning that its group velocity preserves its sign, and thus the spin wave can carry energy only in one of two opposite directions \cite{Lokk_2008}.

In this paper, we investigate the spectra of Damon-Eshbach magnetostatic modes in a dipolar-coupled FM/PM bilayer. A magnetostatic wave propagating in the FM film induces a magnetic response in the PM layer. This response, mediated by stray fields, affects the primary spin wave, thereby modifying its spectrum. Within our model, the PM is characterized by a scalar magnetic susceptibility $\chi$ (or magnetic permeability $\mu = 1 + 4\pi\chi$). We assume that at the Curie temperature $T_C$, the PM undergoes a second-order phase transition associated with the onset of ferromagnetic order. Consequently, its susceptibility above $T_C$ follows the Curie-Weiss law, $\chi = C/(T-T_C)$, where $C$ is the Curie constant. Thus, if the system temperature $T$ is maintained in the vicinity of the PM Curie temperature, its susceptibility can reach very large values, leading to a strong temperature dependence of the spin-wave spectrum. We find that when the magnetic permeability $\mu$ exceeds a critical value $\mu_c$, the MSSW becomes unidirectional. By varying either the system temperature or the external magnetic field, one can switch the unidirectional MSSW regime on and off. These findings could be useful for the development of tunable magnonic logic devices, such as spin-wave diodes \cite{Grassi_2020,Chen_2022}, which transmit energy in only one direction.

The paper is organized as follows. In Sec. \ref{sec:geometry}, we describe the geometry of the system and the governing equations. In Sec. \ref{sec:spectra}, we calculate the spectra of BVMSWs and MSSWs propagating parallel and perpendicular to the FM magnetization, respectively. We determine the condition for the MSSW to become unidirectional and calculate the magnitude of its frequency nonreciprocity. In Sec. \ref{sec:arbitrary}, these results are generalized to the case of an arbitrary propagation direction of magnetostatic waves relative to the FM magnetization, and the dependence of the MSSW cutoff angle on the magnetic permeability $\mu$ is established. Finally, we discuss our results in Sec. \ref{sec:conclusion}.

\section{System Geometry and Basic Equations}
\label{sec:geometry}

\begin{figure}
\includegraphics[width=\columnwidth]{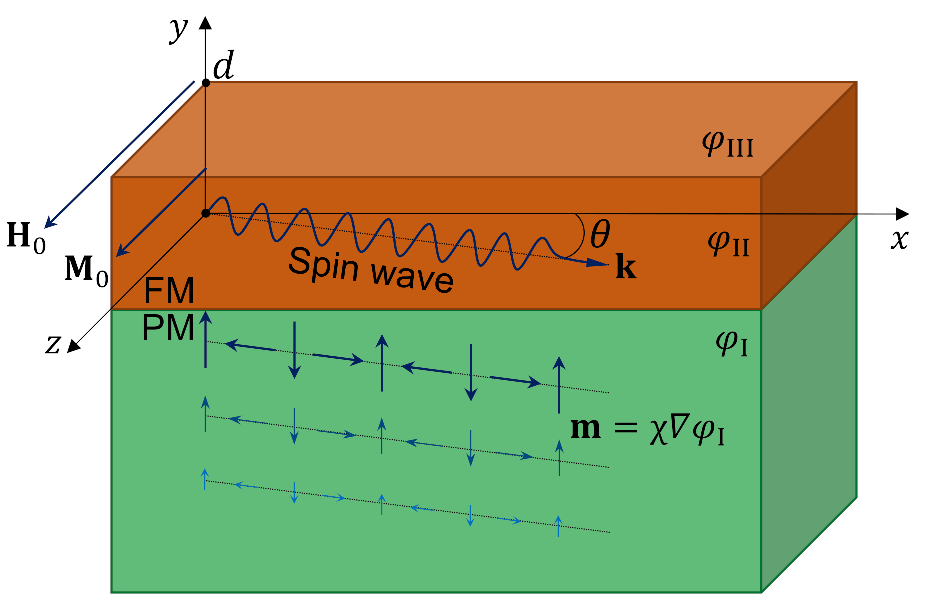} % Замените на имя вашего файла
\caption{Schematic illustration of the ferromagnet/paramagnet (FM/PM) bilayer. The spin wave in the FM film, with wave vector $\mathbf{k}$ at an angle $\theta$ to the $x$ axis, is shown by a wavy arrow. The induced magnetization $\mathbf{m}$ in the PM is depicted by counterclockwise-rotating arrows.}
\label{fig:geometry}
\end{figure}

Consider an FM film of thickness $d$ ($0 < y < d$) placed on a PM half-space ($y < 0$). Let the static uniform magnetization component $\mathbf{M}_0$ of the FM film  lie in its plane and be parallel to the $z$ axis (Fig.~\ref{fig:geometry}). We assume that the system temperature $T$ is close to the Curie temperature $T_C$ of the PM, so that $T \geq T_C$. At the same time, the Curie temperature of the FM film is much higher than $T$, so we can neglect the effect of thermal disordering within it. In the magnetostatic approximation, the distribution of the scalar potential $\varphi(\mathbf{r},t)$ of the magnetic field strength $\mathbf{H}(\mathbf{r},t)$ ($\mathbf{H} = \nabla\varphi$) associated with a spin wave in the FM film can be described by the following equations and corresponding boundary conditions \cite{gurevich_magnetization_1996}:

\begin{subequations}
\begin{align}
\Delta\varphi_{\text{I(III)}} &= 0, \label{eq:laplace_I_III} \\
\mu_1 \left(\frac{\partial^2 \varphi_{\text{II}}}{\partial x^2} + \frac{\partial^2 \varphi_{\text{II}}}{\partial y^2}\right) + \frac{\partial^2 \varphi_{\text{II}}}{\partial z^2} &= 0, \label{eq:laplace_II}
\end{align}
\end{subequations}

\begin{subequations}
\begin{align}
\varphi_{\text{II}}\big|_{y=0} &= \varphi_{\text{I}}\big|_{y=0}, \label{eq:bc_phi_y0} \\
\left(\mu_1 \frac{\partial \varphi_{\text{II}}}{\partial y} - i\mu_2 \frac{\partial \varphi_{\text{II}}}{\partial x}\right)\bigg|_{y=0} &= \mu \frac{\partial \varphi_{\text{I}}}{\partial y}\bigg|_{y=0}, \label{eq:bc_dphi_y0}
\end{align}
\end{subequations}

\begin{subequations}
\begin{align}
\varphi_{\text{II}}\big|_{y=d} &= \varphi_{\text{III}}\big|_{y=d}, \label{eq:bc_phi_yd} \\
\left(\mu_1 \frac{\partial \varphi_{\text{II}}}{\partial y} - i\mu_2 \frac{\partial \varphi_{\text{II}}}{\partial x}\right)\bigg|_{y=d} &= \frac{\partial \varphi_{\text{III}}}{\partial y}\bigg|_{y=d}. \label{eq:bc_dphi_yd}
\end{align}
\end{subequations}
Here $\varphi_{\text{I}}$, $\varphi_{\text{II}}$, and $\varphi_{\text{III}}$ correspond to the regions $y < 0$, $0 < y < d$, and $y > d$, respectively (see Fig.~\ref{fig:geometry}),
\begin{equation}
\mu_1 = \frac{\gamma^2 H_0 (H_0 + 4\pi M_0) - \omega^2}{\gamma^2 H_0^2 - \omega^2}, \quad \mu_2 = \frac{4\pi\gamma M_{0z} \omega}{\gamma^2 H_0^2 - \omega^2},
\label{eq:mu_components}
\end{equation}

\noindent
$i$ is the imaginary unit, $\gamma > 0$ is the gyromagnetic ratio, $\omega$ is the spin-wave frequency, $H_0$ is the external magnetic field, and $M_{0z} = \pm M_0$. The effect of the PM on magnetostatic waves propagating in the FM is described by the magnetic permeability $\mu$ entering the boundary condition (\ref{eq:bc_dphi_y0}), since the magnetic flux density in the region $y < 0$ is $\mu\nabla\varphi_{\text{I}}$, and the PM magnetization is $\mathbf{m} = \chi\nabla\varphi_{\text{I}}$. The case $\mu = 1$ corresponds to the Damon-Eshbach solution \cite{damon_magnetostatic_1961}, while the case $\mu \to \infty$ corresponds to the solution described in Ref.~\cite{Vashkovskii_2006}.

\section{MSSW and BVMSW modes at Special Propagation Directions ($\mathbf{k}\perp\mathbf{M}_0$ and $\mathbf{k}\parallel\mathbf{M}_0$)}
\label{sec:spectra}

\begin{figure}
\includegraphics[width=\columnwidth]{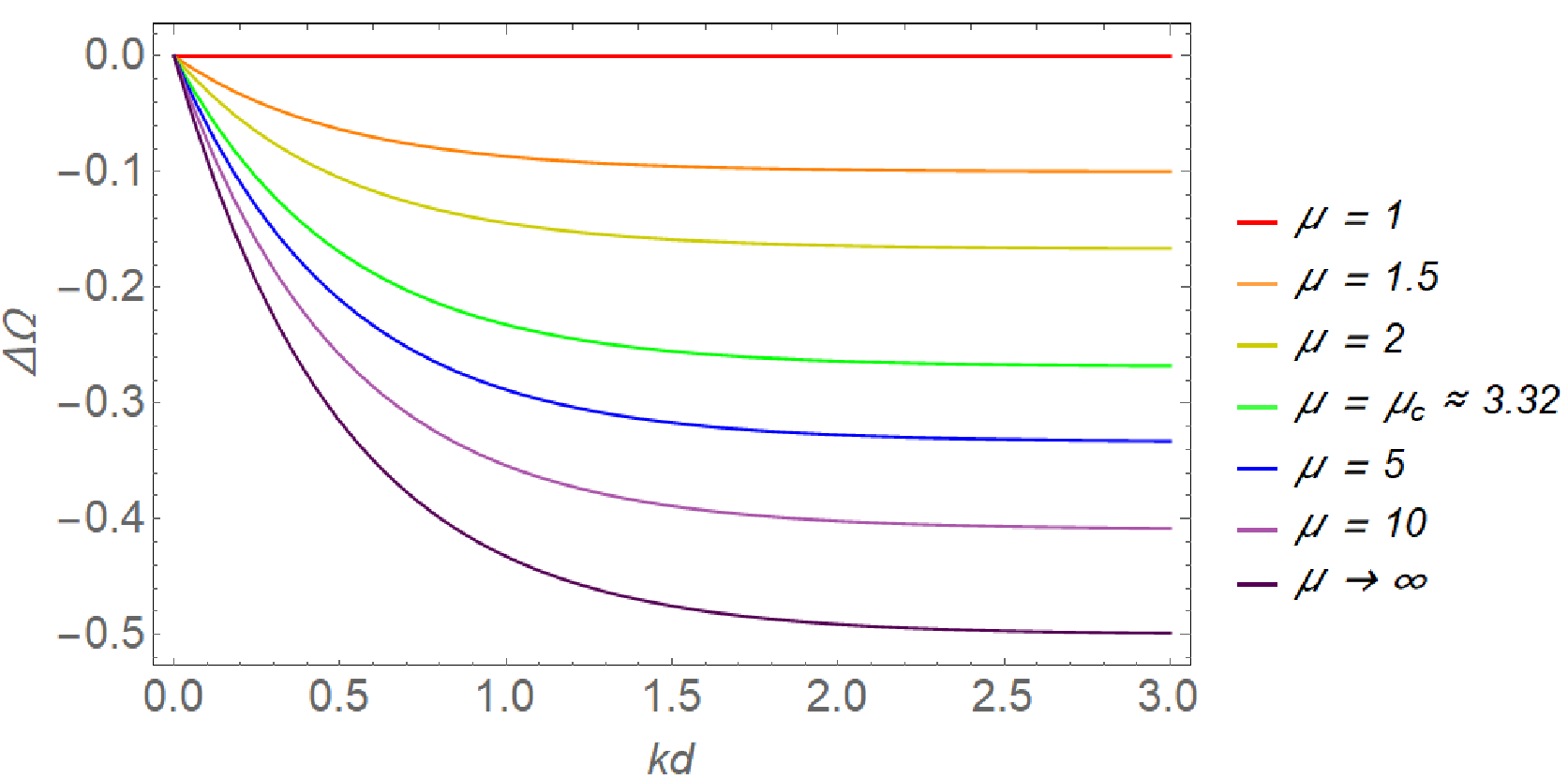}
\caption{Dependencies of the MSSW frequency shift $\Delta\Omega$ on the wave vector modulus $k$ ($\mathbf{k}\perp\mathbf{M}_0$) at different values of the magnetic permeability $\mu$ in the FM/PM bilayer.}
\label{fig:MSSW_shift}
\end{figure}

\begin{figure*}
\includegraphics[width=\textwidth]{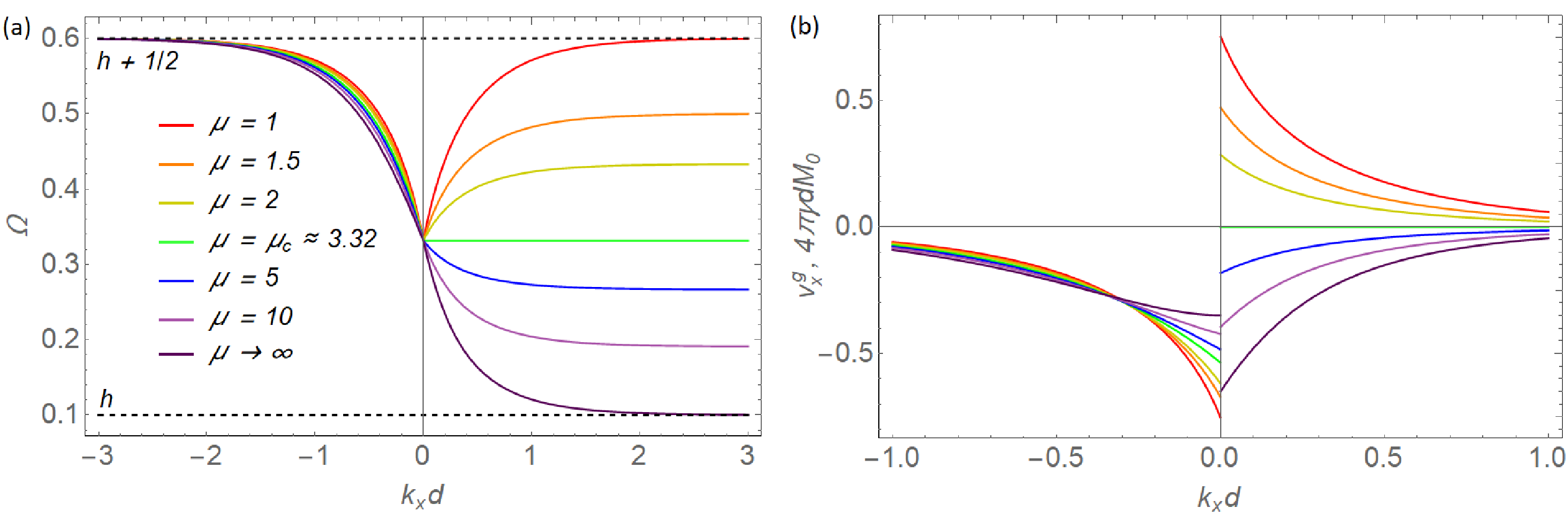}
\caption{Dependencies of (a) frequency $\Omega$ and (b) group velocity $v_x^g$ of MSSWs on the $x$ component of the wave vector $k_x$ ($\mathbf{k}\perp\mathbf{M}_0$) at different values of the magnetic permeability $\mu$ in the FM/PM bilayer. For the calculations, $h = 0.1$ ($\mu_c \approx 3.32$) and $s_M = 1$ were adopted.}
\label{fig:MSSW_omega_vg}
\end{figure*}

We first determine the spectrum of MSSWs propagating in the FM film plane perpendicular to its magnetization (parallel to the $x$ axis). We seek solutions of Eqs.~(\ref{eq:laplace_I_III}) and (\ref{eq:laplace_II}) in the form
\begin{subequations}
\begin{align}
\varphi_{\text{I}} &= a_1 e^{i\omega t} e^{-ik_x x} e^{ky}, \label{eq:phi_I_MSSW} \\
\varphi_{\text{II}} &= e^{i\omega t} e^{-ik_x x} (a_2 e^{-ky} + a_3 e^{ky}), \label{eq:phi_II_MSSW} \\
\varphi_{\text{III}} &= a_4 e^{i\omega t} e^{-ik_x x} e^{-k(y-d)}. \label{eq:phi_III_MSSW}
\end{align}
\end{subequations}
Here $k_x$ is the $x$ component of the spin-wave vector $\mathbf{k}$ (in this case $|k_x| = k$). Substituting the potentials (\ref{eq:phi_I_MSSW})--(\ref{eq:phi_III_MSSW}) into the boundary conditions (\ref{eq:bc_phi_y0}), (\ref{eq:bc_dphi_y0}), (\ref{eq:bc_phi_yd}), and (\ref{eq:bc_dphi_yd}), we obtain a system of equations for the constants $a_1$, $a_2$, $a_3$, and $a_4$. From the condition for the existence of a nontrivial solution of this system, we find the MSSW spectrum in terms of the dimensionless frequency $\Omega = \omega/(4\pi\gamma M_0)$:
\begin{equation}
\Omega = \frac{\Delta\Omega}{2}s + \sqrt{\left(h + \frac{1+\Delta\Omega}{2}\right)^2 - \frac{1}{4}\frac{1-\tanh kd}{1+\tanh kd}},
\label{eq:MSSW_spectrum}
\end{equation}
\begin{equation}
\Delta\Omega \equiv \Omega(s=1) - \Omega(s=-1) = -\frac{\mu-1}{\mu+1}\frac{\tanh kd}{1+\tanh kd},
\label{eq:MSSW_nonreciprocity}
\end{equation}
where we have introduced the notation $h = H_0/(4\pi M_0)$ and $s = (k_x/k)s_M$, with $s_M = M_{0z}/M_0 = \pm 1$. At $k = 0$, we have $\Omega = \sqrt{h(h+1)}$, while at $kd \gg 1$, $\Omega(s=-1) = h + 1/2$ and $\Omega(s=1) = h + 1/(\mu+1)$. In this case, if $s = 1$, the MSSW is localized near the FM/PM interface, while if $s = -1$, it is localized near the FM/vacuum interface. For the case of an FM film bounded by vacuum on both sides ($\mu = 1$), Eq.~(\ref{eq:MSSW_spectrum}) yields the well-known result for the Damon-Eshbach surface mode \cite{damon_magnetostatic_1961}. In contrast, in the presence of a PM ($\mu \neq 1$), the spin-wave spectrum becomes nonreciprocal, i.e., dependent on the sign of $k_x$ or $M_{0z}$, so that $\Delta\Omega \neq 0$ (Fig.~\ref{fig:MSSW_shift}). The magnitude of this nonreciprocity reaches its maximum absolute value at $\mu \to \infty$ ($T = T_C$). Note that the sign of the nonreciprocal term in the FM/PM differs from the sign of this term in FM/normal metal \cite{seshadri_surface_1970} or FM/superconductor \cite{Kuznetsov_2022}. Figure~\ref{fig:MSSW_omega_vg}(a) shows the dependencies of $\Omega$ on $k_x$ for various $\mu$. We see that at $\mu > \mu_c = \sqrt{1+1/h}$, the MSSWs become backward (in the region $h < \Omega < \sqrt{h(h+1)}$) and unidirectional (throughout the entire existence region). The former means that the group $v_x^g = \partial\omega/\partial k_x$ and phase $v_x^p = \omega/k_x$ velocities have opposite signs, while the latter means that $v_x^g$ preserves its sign, i.e., such a spin wave can carry energy in only one direction. For clarity, Fig.~\ref{fig:MSSW_omega_vg}(b) shows the dependencies of $v_x^g$ on $k_x$ for various $\mu$.

\begin{figure}
\includegraphics[width=\columnwidth]{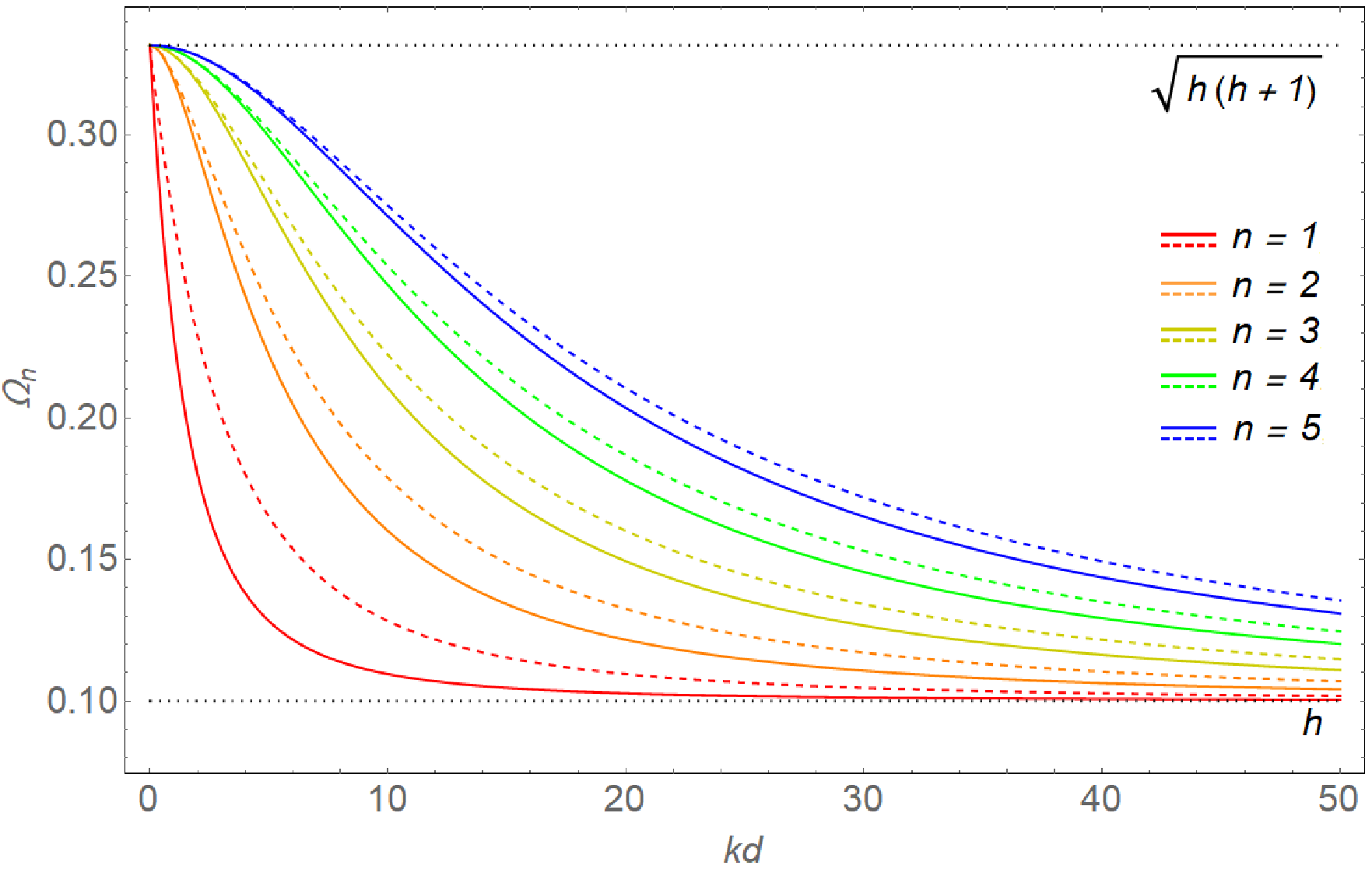}
\caption{Frequencies of the first five BVMSW modes $\Omega_n$ as functions of the wave vector modulus $k$ ($\mathbf{k}\parallel\mathbf{M}_0$) in the FM/PM bilayer. The solid lines correspond to the case $\mu \to \infty$, while the dashed lines correspond to the case $\mu = 1$. For the calculations, $h = 0.1$ was adopted.}
\label{fig:BVMSW_spectrum}
\end{figure}

We now determine the spectrum of BVMSWs propagating in the FM film plane parallel to its magnetization. We seek solutions of Eqs.~(\ref{eq:laplace_I_III}) and (\ref{eq:laplace_II}) in the form
\begin{subequations}
\begin{align}
\varphi_{\text{I}} &= b_1 e^{i\omega t} e^{-ik_z z} e^{ky}, \label{eq:phi_I_BVMSW} \\
\varphi_{\text{II}} &= e^{i\omega t} e^{-ik_z z} (b_2 e^{-i\alpha ky} + b_3 e^{i\alpha ky}), \label{eq:phi_II_BVMSW} \\
\varphi_{\text{III}} &= b_4 e^{i\omega t} e^{-ik_z z} e^{-k(y-d)}. \label{eq:phi_III_BVMSW}
\end{align}
\end{subequations}
Here $k_z$ is the $z$ component of the spin-wave vector $\mathbf{k}$ (in this case $|k_z| = k$). Substituting $\varphi_{\text{II}}$ into Eq.~(\ref{eq:laplace_II}), we find $\alpha = 1/\sqrt{-\mu_1}$, which implies $\mu_1 < 0$. From the condition for the existence of a nontrivial solution of the system of equations for the constants $b_1$, $b_2$, $b_3$, and $b_4$, which arises after substituting the potentials (\ref{eq:phi_I_BVMSW})--(\ref{eq:phi_III_BVMSW}) into the boundary conditions (\ref{eq:bc_phi_y0}), (\ref{eq:bc_dphi_y0}), (\ref{eq:bc_phi_yd}), and (\ref{eq:bc_dphi_yd}), we find the BVMSW spectrum in terms of the dimensionless frequency $\Omega_n = \omega_n/(4\pi\gamma M_0)$:
\begin{equation}
\Omega_n = \sqrt{h\left(h + \frac{\alpha_n^2}{1+\alpha_n^2}\right)}, \quad n = 1, 2, 3, \dots,
\label{eq:BVMSW_spectrum}
\end{equation}
where the index $n$ denotes the mode number of the bulk wave, and $\alpha_n$ is a solution of the equation
\begin{equation}
\begin{split}
\cot(\alpha_n kd) &= \frac{1}{\mu+1}\left(\mu\alpha_n - \frac{1}{\alpha_n}\right), \\
\frac{\pi(n-1)}{kd} &\leq \alpha_n \leq \frac{\pi n}{kd}.
\end{split}
\label{eq:BVMSW_alpha}
\end{equation}
At $k = 0$, we have $\alpha_n \to \infty$ and $\Omega_n = \sqrt{h(h+1)}$, while at $k \to \infty$, $\alpha_n = 0$ and $\Omega_n = h$. Figure~\ref{fig:BVMSW_spectrum} shows the spectra of the first five BVMSW modes $\Omega_n$ as functions of $k$, obtained numerically. The solid lines correspond to the case $\mu \to \infty$, while the dashed lines correspond to the case of Damon-Eshbach bulk waves ($\mu = 1$) \cite{damon_magnetostatic_1961}. Note that the spectrum does not depend on the sign of $k_z$, i.e., in this configuration the spin waves are reciprocal.

Thus, the PM half-space has a significant effect on the MSSW  spectrum ($\mathbf{k}\perp\mathbf{M}_0$) and a relatively weak effect on the BVMSW spectrum ($\mathbf{k}\parallel\mathbf{M}_0$). In addition to the emergence of unidirectionality and backward-wave properties, at $\mu > \mu_c$ the frequency ranges of MSSWs ($h < \Omega < h + 1/2$) and BVMSWs ($h \leq \Omega_n \leq \sqrt{h(h+1)}$) overlap. This situation is possible because MSSWs and BVMSWs propagate in different directions in this case. We now investigate what happens when both MSSWs and BVMSWs propagate along the same arbitrary direction relative to the magnetization $\mathbf{M}_0$.

\section{MSSW AND BVMSW modes FOR ARBITRARY PROPAGATION DIRECTION}
\label{sec:arbitrary}
\begin{figure*}
\includegraphics[width=\textwidth]{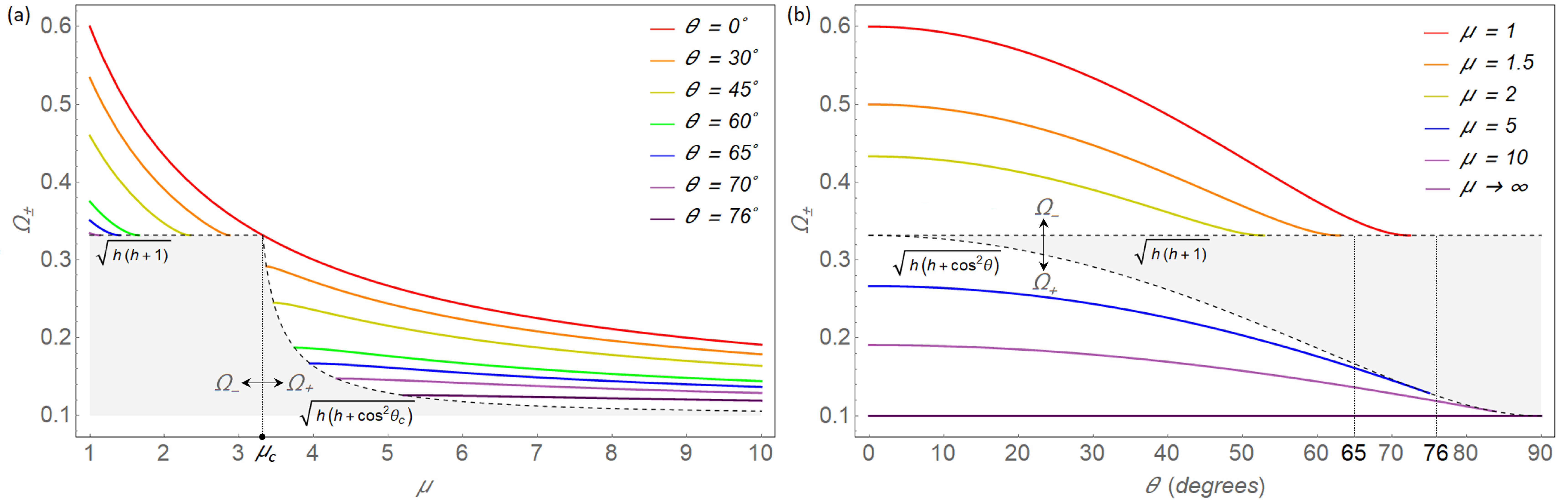}
\caption{Dependencies of the MSSW  frequencies $\Omega_+$ and $\Omega_-$ on (a) magnetic permeability $\mu$ and (b) propagation angle $\theta$ in the FM/PM bilayer for the case $kd \gg 1$. The light gray shaded area corresponds to the forbidden region for MSSWs. The double arrows indicate the frequency ranges of $\Omega_+$ and $\Omega_-$. For the calculations, $h = 0.1$ ($\mu_c \approx 3.32$) and $s_M = 1$ were adopted.}
\label{fig:Omega_mu_theta}
\end{figure*}

\begin{figure}
\includegraphics[width=\columnwidth]{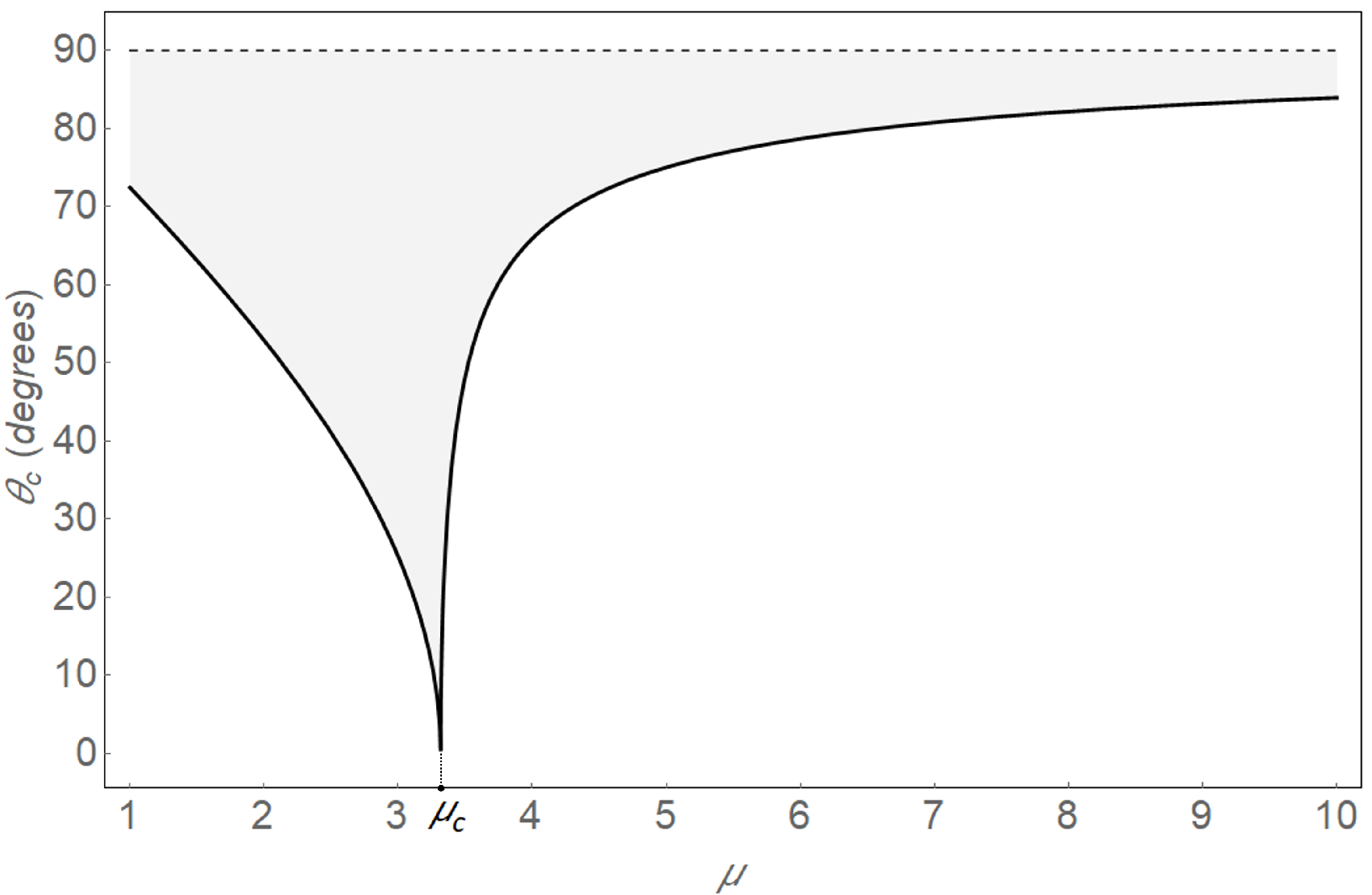}
\caption{Dependence of the cutoff angle $\theta_c$ on the magnetic permeability $\mu$ in the FM/PM bilayer. The light gray shaded area corresponds to the forbidden region for MSSWs in the case $s_M=1$. For the calculations, $h = 0.1$ ($\mu_c \approx 3.32$) was adopted.}
\label{fig:theta_c}
\end{figure}

We first consider MSSWs propagating in an arbitrary direction relative to the magnetization $\mathbf{M}_0$ in the case of a thick FM film ($kd \gg 1$). Focusing on the magnetostatic wave localized at the FM/PM interface, we seek solutions of Eqs.~(\ref{eq:laplace_I_III}) and (\ref{eq:laplace_II}) in the form
\begin{subequations}
\begin{align}
\varphi_{\text{I}} &= \varphi_0 e^{i\omega t} e^{-i\mathbf{k}\cdot\mathbf{r}} e^{ky}, \label{eq:phi_I_arb} \\
\varphi_{\text{II}} &= \varphi_0 e^{i\omega t} e^{-i\mathbf{k}\cdot\mathbf{r}} e^{-\kappa y}, \label{eq:phi_II_arb}
\end{align}
\end{subequations}
where $\varphi_0$ is an arbitrary constant, $\mathbf{k} = k(\cos\theta, 0, \sin\theta)$, and $\theta$ is the angle between the wave vector $\mathbf{k}$ and the $x$ axis (see Fig.~\ref{fig:geometry}). Substituting $\varphi_{\text{II}}$ into Eq.~(\ref{eq:laplace_II}), we find $\kappa = k\sqrt{\cos^2\theta + \sin^2\theta/\mu_1}$. The condition for $\kappa$ to be real implies that MSSWs can propagate in the regions $\Omega > \sqrt{h(h+1)}$ and $\Omega < \sqrt{h(h+\cos^2\theta)}$. In the region $\sqrt{h(h+\cos^2\theta)} \leq \Omega \leq \sqrt{h(h+1)}$, the parameter $\kappa$ becomes imaginary or zero (at the region boundaries), which cannot correspond to surface waves. From the boundary condition (\ref{eq:bc_dphi_y0}), we obtain the dispersion equation
\begin{equation}
\mu_2 k \cos\theta + \mu_1 \kappa = -k\mu,
\label{eq:dispersion_arb}
\end{equation}
the solutions of which have the following form:
\begin{equation}
\begin{split}
\Omega_\pm &= \frac{\omega_\pm}{4\pi\gamma M_0} = \frac{\mu s_M \cos\theta}{\mu^2-1} \\
&\quad \pm \sqrt{\left(\frac{1}{\mu^2-1}-h\right)\left(\frac{\cos^2\theta}{\mu^2-1}-h\right)}.
\end{split}
\label{eq:Omega_pm}
\end{equation}
In Eq.~(\ref{eq:Omega_pm}), the branches $\Omega_+$ and $\Omega_-$ correspond to the regimes $\mu \geq \mu_c$ and $\mu \leq \sqrt{1+\cos^2\theta/h} \leq \mu_c$, respectively. Therefore, at $\mu = 1$, the branch $\Omega_+$ is absent, while the branch $\Omega_-$ reduces to the well-known result for the Damon-Eshbach surface mode (see, e.g., Ref.~\cite{camley_nonreciprocal_1987}), i.e.,
\begin{equation}
\Omega_- = \frac{s_M}{2}\left(\frac{h}{\cos\theta} + (h+1)\cos\theta\right).
\label{eq:Omega_DE_arb}
\end{equation}
Note that the spectra (\ref{eq:Omega_pm}) allow MSSW propagation only in directions for which $s = s_M \cos\theta > 0$. Moreover, the obtained solutions must ensure $\kappa > 0$. From Eq.~(\ref{eq:dispersion_arb}), we then have the additional condition $(\mu + \mu_2 \cos\theta)/\mu_1 < 0$, which implies that the branches $\Omega_+$ and $\Omega_-$ exist in the regions $\mu > \mu_c^+$ (with $\mu_c^+ \geq \mu_c$) and $\mu < \mu_c^-$ (with $\mu_c^- \leq \sqrt{1+\cos^2\theta/h}$), respectively, where
\begin{equation}
\mu_c^+ = \frac{s_M}{\cos\theta}\sqrt{1+\frac{\cos^2\theta}{h}}, \quad \mu_c^- = s_M \cos\theta\sqrt{1+\frac{1}{h}}.
\label{eq:mu_c_pm}
\end{equation}
Thus, in the MSSW spectrum propagating in an arbitrary direction relative to the magnetization $\mathbf{M}_0$, a forbidden region of magnetic permeability $\mu_c^- \leq \mu \leq \mu_c^+$ emerges. In this region, the surface wave does not exist. At $\theta = 0$, the forbidden region disappears, and $\Omega_\pm = h + 1/(\mu+1)$, which coincides with Eq.~(\ref{eq:MSSW_spectrum}) at $kd \gg 1$ and $s = 1$. Instead of the forbidden region of magnetic permeability, one can speak of forbidden angles for magnetostatic wave propagation by introducing the concept of the cutoff angle $\theta_c$ ($0 \leq \theta_c \leq \pi/2$), such that the MSSW can propagate only in the directions $-\theta_c < \theta < \theta_c$ ($s_M = 1$) or $\pi - \theta_c < \theta < \pi + \theta_c$ ($s_M = -1$). The expression for $\cos\theta_c$ has the following form:
\begin{equation}
\cos\theta_c = \begin{cases}
\mu/\mu_c, & \mu \leq \mu_c, \\
\sqrt{1/(\mu^2-\mu_c^2+1)}, & \mu \geq \mu_c.
\end{cases}
\label{eq:cos_theta_c}
\end{equation}
Figure~\ref{fig:Omega_mu_theta} shows the dependencies of $\Omega_\pm$ on $\mu$ at various $\theta$ [Fig.~\ref{fig:Omega_mu_theta}(a)] and the dependencies of $\Omega_\pm$ on $\theta$ at various $\mu$ [Fig.~\ref{fig:Omega_mu_theta}(b)]. The light gray regions indicate the forbidden values of $\mu$ and $\theta$ where MSSWs cannot propagate. The curves $\Omega_+$ and $\Omega_-$ reach the forbidden regions at $\mu = \mu_c^\pm$ or $\theta = \theta_c$ and terminate there. The branch $\Omega_-$ at $\mu = \mu_c^-$ reaches a local minimum equal to $\sqrt{h(h+1)}$, while the branch $\Omega_+$ at $\mu = \mu_c^+$ reaches a local maximum equal to $\sqrt{h(h+\cos^2\theta)}$. Figure~\ref{fig:theta_c} shows the dependence of the cutoff angle $\theta_c$ on the magnetic permeability $\mu$. As can be seen, initially with increasing $\mu$ ($\mu \leq \mu_c$), $\theta_c$ decreases to zero, then at $\mu \geq \mu_c$, it increases to $\pi/2$. Thus, if $s_M = 1$, at $\mu = \mu_c$, MSSW propagation is possible only along the $x$ axis, i.e., at $\theta = 0$, while at $\mu \to \infty$, it is possible in the entire right half-plane, i.e., at $-\pi/2 < \theta < \pi/2$.

We now consider MSSWs and BVMSWs propagating in an FM film of arbitrary thickness $d$ in an arbitrary direction. The potentials are then sought in the following form:
\begin{subequations}
\begin{align}
\varphi_{\text{I}} &= c_1 e^{i\omega t} e^{-i\mathbf{k}\cdot\mathbf{r}} e^{ky}, \label{eq:phi_I_arb_thick} \\
\varphi_{\text{II}} &= e^{i\omega t} e^{-i\mathbf{k}\cdot\mathbf{r}} (c_2 e^{-\kappa_1 y} + c_3 e^{\kappa_2 y}), \label{eq:phi_II_arb_thick} \\
\varphi_{\text{III}} &= c_4 e^{i\omega t} e^{-i\mathbf{k}\cdot\mathbf{r}} e^{-k(y-d)}, \label{eq:phi_III_arb_thick}
\end{align}
\end{subequations}
where the constants $\kappa_{1,2}$ can be either real (MSSW case) or imaginary (BVMSW case). Substituting the potentials into the boundary conditions (\ref{eq:bc_phi_y0}), (\ref{eq:bc_dphi_y0}), (\ref{eq:bc_phi_yd}), and (\ref{eq:bc_dphi_yd}) and requiring a nontrivial solution for $c_1$, $c_2$, $c_3$, and $c_4$, we obtain the dispersion expressions for the MSSW  spectrum
\begin{equation}
\tanh(\kappa d) = \frac{(1+\mu)\mu_1 k\kappa}{k^2(\mu_2^2 \cos^2\theta - \mu - (1-\mu)\mu_2 \cos\theta) - \mu_1^2 \kappa^2}
\label{eq:MSSW_arb_thick}
\end{equation}
and the BVMSW spectrum
\begin{equation}
\Omega_n = \sqrt{h\left(h + \frac{\cos^2\theta + \alpha_n^2}{1+\alpha_n^2}\right)}, \quad n = 1, 2, 3, \dots,
\label{eq:BVMSW_arb_thick}
\end{equation}
where $\alpha_n$ ($\alpha_n^2 = -\kappa^2/k^2$) is a solution of the equation
\begin{equation}
\begin{split}
\cot(\alpha_n kd) &= \frac{\mu_1^2 \alpha_n^2 - \mu - (1-\mu)\mu_2 \cos\theta + \mu_2^2 \cos^2\theta}{(1+\mu)\mu_1 \alpha_n}, \\ \frac{\pi(n-1)}{kd} &\leq \alpha_n \leq \frac{\pi n}{kd},
\end{split}
\label{eq:BVMSW_alpha_arb}
\end{equation}
in which $\mu_1$ and $\mu_2$ must be expressed in terms of $\alpha_n$, i.e.,
\begin{subequations}
\begin{align}
\mu_1(\alpha_n) &= -\frac{\sin^2\theta}{\alpha_n^2+\cos^2\theta}, \label{eq:mu1_alpha} \\
\mu_2(\alpha_n) &= -s_M \sqrt{\frac{\alpha_n^2+1}{\alpha_n^2+\cos^2\theta}\left(\frac{\alpha_n^2+1}{\alpha_n^2+\cos^2\theta}+\frac{1}{h}\right)}. \label{eq:mu2_alpha}
\end{align}
\end{subequations}
\begin{figure*}
\includegraphics[width=\textwidth]{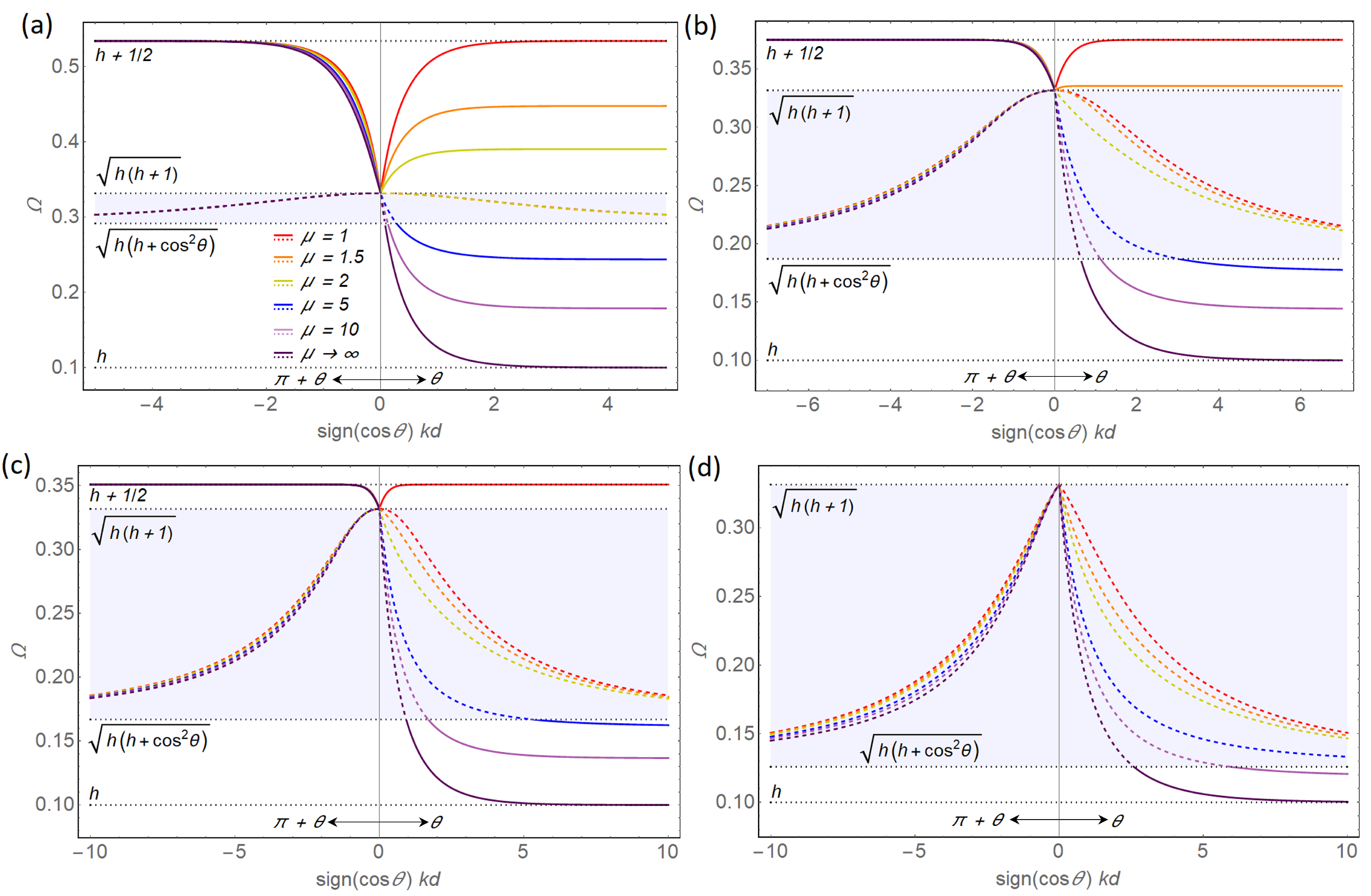}
\caption{Spectra of MSSWs (solid lines) and the first BVMSW mode (dashed lines) at various values of the magnetic permeability $\mu$ and (a) $\theta = 30^\circ$, (b) $\theta = 60^\circ$, (c) $\theta = 65^\circ$, and (d) $\theta = 76^\circ$ in the FM/PM bilayer. The light blue shaded area corresponds to the BVMSW region. The double arrows indicate the forward ($\theta$) and backward ($\pi+\theta$) propagation directions. For the calculations, $h = 0.1$ ($\mu_c \approx 3.32$) and $s_M = 1$ were adopted.}
\label{fig:MSSW_BVMSW_arb}
\end{figure*}
\begin{figure*}
\includegraphics[width=\textwidth]{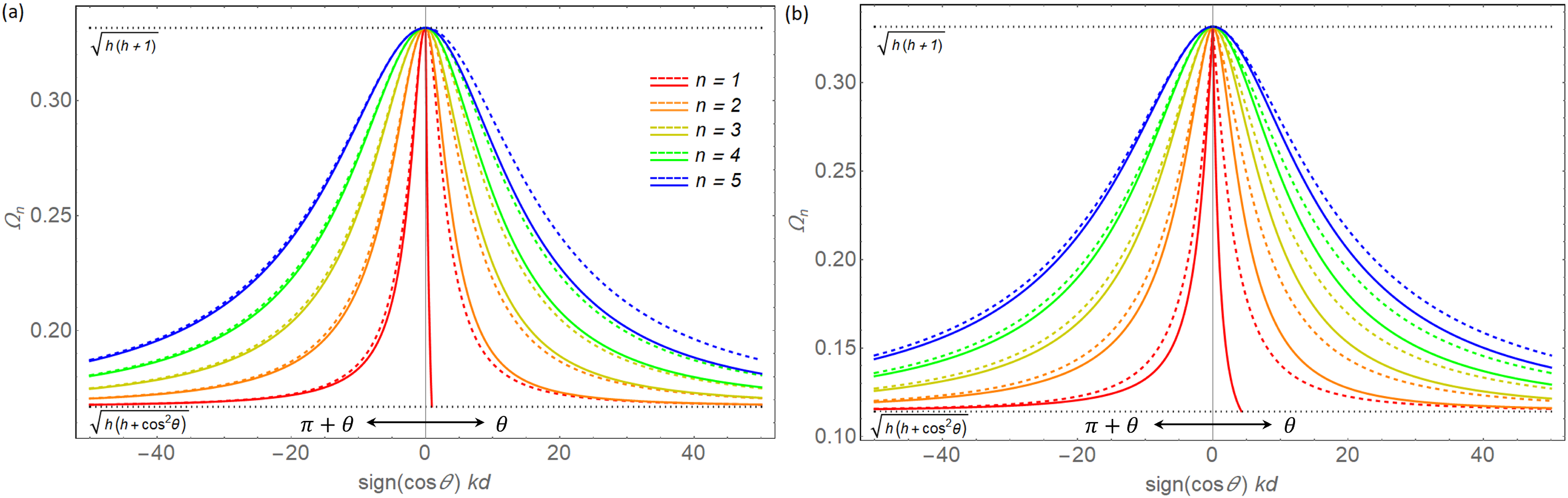}
\caption{Frequencies of the first five BVMSW modes $\Omega_n$ as functions of the wave vector modulus $k$ in the FM/PM bilayer at (a) $\theta = 65^\circ$ and (b) $\theta = 80^\circ$. The solid lines correspond to the case $\mu \to \infty$, while the dashed lines correspond to the case $\mu = 1$. The double arrows indicate the forward ($\theta$) and backward ($\pi+\theta$) propagation directions. For the calculations, $h = 0.1$ and $s_M = 1$ were adopted.}
\label{fig:BVMSW_modes_arb}
\end{figure*}

\noindent Figure~\ref{fig:MSSW_BVMSW_arb} shows the MSSW (solid lines) and first BVMSW mode (dashed lines) spectra, obtained numerically at various $\mu$ and $\theta$. The light blue shading indicates the BVMSW propagation region ($\sqrt{h(h+\cos^2\theta)} \leq \Omega_n \leq \sqrt{h(h+1)}$). At $\theta = 0$, this region is absent, and only MSSW propagation is allowed ($h < \Omega < h+1/2$, see Fig.~\ref{fig:MSSW_omega_vg}(a)). When $\theta \neq 0$ and $|\theta| < \theta_c$ for any $\mu$, the MSSW region is split into two parts by the BVMSW region [Fig.~\ref{fig:MSSW_BVMSW_arb}(a)]. In this case, the first BVMSW mode spectrum, as $k$ increases, reaches the boundary of its region and transitions into the MSSW spectrum without a derivative discontinuity. Since this is accompanied by the transformation of bulk waves into surface ones, at the boundary frequency $\sqrt{h(h+\cos^2\theta)}$, magnetostatic waves exist with an amplitude constant across the film thickness \cite{Vashkovskii_2006}. It should be noted that, within the exchange-free magnetostatic approximation, hybridization between bulk and surface modes does not occur. If exchange were included, these modes would hybridize, and an avoided crossing might appear in their spectra instead of a sharp transition~\cite{De_Wames_2}. With further increase of the angle $\theta$, the inequality $|\theta| < \theta_c$ is violated for some values of $\mu$, and the propagation of the corresponding MSSW is forbidden. Thus, in Fig.~\ref{fig:MSSW_BVMSW_arb}(b), the MSSW with $\mu = 2$ is absent; in Fig.~\ref{fig:MSSW_BVMSW_arb}(c), with $\mu = 1.5$ and $2$; and in Fig.~\ref{fig:MSSW_BVMSW_arb}(d), with $\mu = 1$, $1.5$, $2$, and $5$ (note the vertical dotted lines in Fig.~\ref{fig:Omega_mu_theta}(b)). The spectra of the first five BVMSW modes for the cases $\mu = 1$ (dashed lines) and $\mu \to \infty$ (solid lines) are shown separately in Fig.~\ref{fig:BVMSW_modes_arb}. As can be seen, only the first BVMSW mode transitions into the MSSW, while for the other modes, the boundary frequency $\sqrt{h(h+\cos^2\theta)}$ is an asymptote. In contrast to the case $\theta = \pi/2$, the BVMSW spectrum becomes nonreciprocal, i.e., $\Omega_n(\theta) - \Omega_n(\pi+\theta) \neq 0$.

Thus, we have found that for the same arbitrary propagation direction of MSSWs and BVMSWs, their spectra do not cross. Instead, the MSSW region is split into two parts by the BVMSW one, and the surface wave and the first bulk mode can transition into each other when reaching the boundary frequency $\sqrt{h(h+\cos^2\theta)}$.

\section{Conclusion}
\label{sec:conclusion}

In this work, we have calculated the spectra of magnetostatic waves (MSSWs and BVMSWs) in an in-plane magnetized FM/PM bilayer, where the FM and PM layers are coupled by dipolar interaction. We have shown that MSSWs propagating perpendicular to the static magnetization component $\mathbf{M}_0$ of the FM become unidirectional if the magnetic permeability $\mu$ of the PM exceeds the critical value $\mu_c$. For unidirectional waves, the sign of the group velocity is preserved, and therefore they can carry energy in only one direction. Note that the direction of energy transport can be reversed by inverting the external magnetic field. In addition, at $\mu > \mu_c$ in the frequency range $h < \Omega < \sqrt{h(h+1)}$, the MSSW becomes backward. This property makes the FM/PM bilayer similar to so-called left-handed media \cite{electrodynamics_1968}, characterized by negative dielectric and magnetic permeabilities, in which the Doppler \cite{Chumak_2010} and Vavilov-Cherenkov \cite{Pafomov_1959} effects are reversed, and other unusual effects are observed. We have calculated the MSSW mode frequency nonreciprocity $\Delta\Omega$ and shown that it reaches its maximum (in absolute value) at $\mu \to \infty$. In the case of an arbitrary propagation direction of magnetostatic waves relative to $\mathbf{M}_0$, in addition to the aforementioned effects, the MSSW spectrum splits into two regions ($\sqrt{h(h+1)} < \Omega < h + 1/2$ and $h < \Omega < \sqrt{h(h+\cos^2\theta)}$), between which the BVMSW region ($\sqrt{h(h+\cos^2\theta)} \leq \Omega_n \leq \sqrt{h(h+1)}$) is located. Moreover, within the exchange-free magnetostatic approximation, the MSSW and the first BVMSW mode transition into each other at the boundary frequency $\sqrt{h(h+\cos^2\theta)}$. At $\theta \neq 0$, the MSSWs become unidirectional when $\mu > \mu_c^+ > \mu_c$. In addition, in the range $\mu_c^- \leq \mu \leq \mu_c^+$ of magnetic permeability, the propagation of MSSWs along the FM/PM interface is forbidden.

\begin{figure}[b]
\includegraphics[width=\columnwidth]{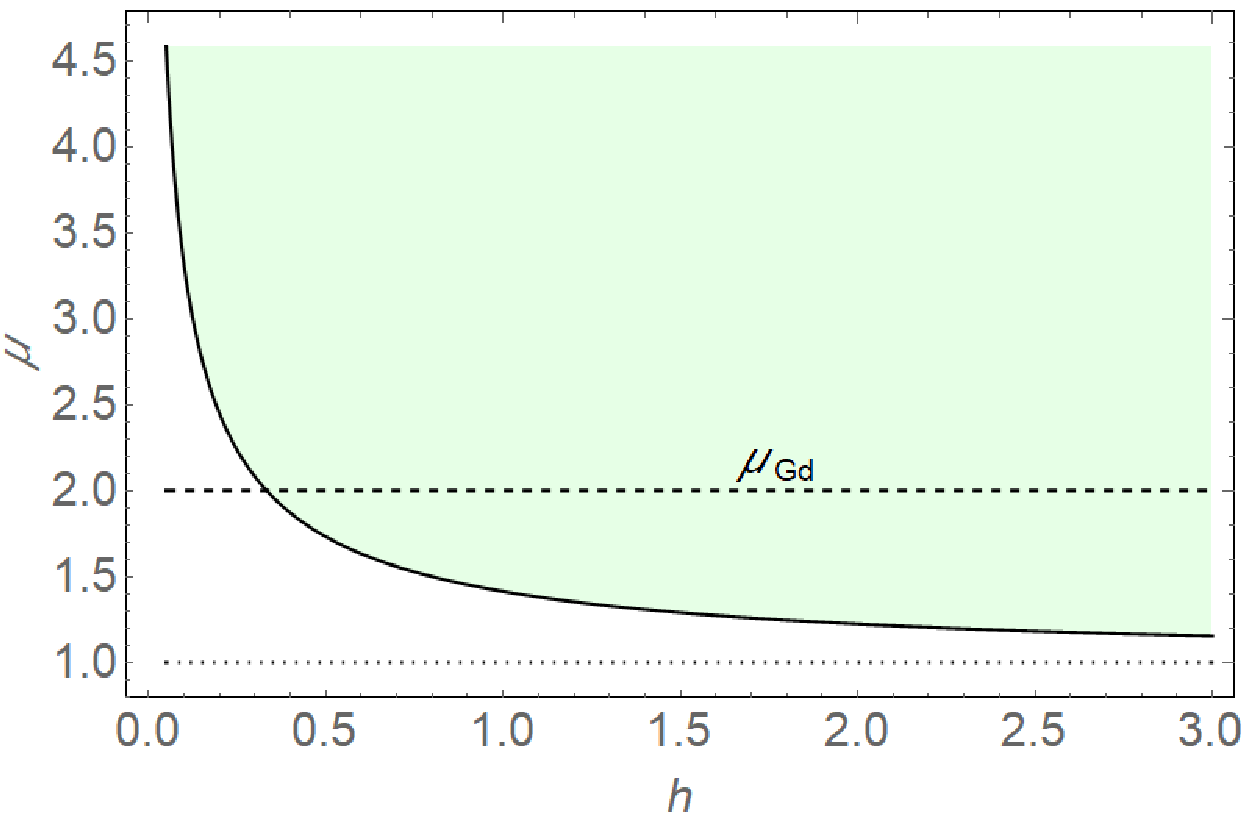}
\caption{Region of unidirectional waves in the parameter space of $\mu$ and $h$ (shaded in light green). The dashed line shows the value of the magnetic permeability of Gd ($\mu_{\text{Gd}}$) corresponding to room temperature ($T = 298$~K).}
\label{fig:unidirectional_region}
\end{figure}

If the PM undergoes a second-order phase transition associated with the onset of ferromagnetic ordering, then in the vicinity of the Curie temperature $T_C$ of such a transition, the magnetic permeability $\mu$ can reach arbitrarily large values, while the PM magnetization remains finite \cite{Kuznetsov_2022}. Then, by varying the system temperature under constant external magnetic field conditions, one can switch on and off the unidirectional wave regime. On the other hand, since the critical permeability $\mu_c$ (or $\mu_c^+$) depends on the external magnetic field, switching on and off this regime can be achieved by varying the magnitude of the external field at constant temperature. Figure~\ref{fig:unidirectional_region} shows the boundary of the unidirectional MSSW region in the parameter space of $\mu$ and $h$ for the case $\mathbf{k} \perp \mathbf{M}_0$. It can be seen that as the external field $h$ increases, the value of $\mu$ required for the transition to the unidirectional wave region decreases. Therefore, for such a transition, it is sufficient to have a PM material whose magnetic permeability at a given temperature is moderately larger than unity. In particular, Gd ($T_C \approx 293$~K) can be considered as such a material, whose magnetic permeability $\mu_{\text{Gd}}$ at room temperature ($T = 298$~K) is approximately equal to 2 (see Fig.~\ref{fig:unidirectional_region}). For other temperature ranges, the solid solution Ni$_x$Cu$_{1-x}$, whose Curie temperature strongly depends on the Cu concentration \cite{ahern_spontaneous_1958}, can be used as a PM material, for example. Thus, the key is to select a PM material whose $T_C$ lies within the intended operating temperature range, while the Curie constant $C$ determines the temperature window where $\mu$ exceeds unity.

We now discuss the range of parameters in which the model under consideration is valid. We assume that the exchange interaction in the FM film can be neglected, which is valid when $kL\ll1$, where $L=\sqrt{2A/M_0^2}$ is the exchange length, and $A$ is the exchange stiffness. Furthermore, many PMs are normal metals (including the proposed Gd and Ni$_x$Cu$_{1-x}$). This means that the dipolar coupling between the spin wave and eddy currents can also contribute to spectral modifications~\cite{seshadri_surface_1970,De_Wames_1970,Mruczkiewicz_2014}. However, this contribution will be negligible if the spin wavelength is much smaller than the skin depth $\delta$ of the PM, i.e., $k\delta\gg1$. These conditions yield the following range for the wave vector $k$ where the model is valid: $1/\delta\ll k\ll1/L$. At typical magnonic frequencies $\omega\sim10^9-10^{10}\text{ s}^{-1}$, for highly conductive metals the skin depth is $\delta\sim10^{-4}\text{ cm}$, while the exchange length is $L\sim10^{-7}-10^{-6}\text{ cm}$. This gives the applicable range $10^4\text{ cm}^{-1}\ll k\ll10^6-10^7\text{ cm}^{-1}$, which corresponds to spin wavelengths from hundreds of nanometers to several micrometers and covers the typical operating regime of magnonic devices. We also assume that the magnetic susceptibility of the PM is purely real and neglect the imaginary part, which is valid when $\omega\tau\ll1$, where $\tau$ is a relaxation time~\cite{Balanda_2013}. As the system approaches $T_C$, the relaxation time increases according to $\tau\propto (T-T_C)^{-\nu}$, where $\nu > 0$. However, ultrasonic propagation 
experiments show that for Gd, even very close to $T_C$ ($(T-T_C)/T_C = 3 \times 10^{-3}$), the relaxation time remains relatively small ($\tau \sim 10^{-10}$ s) \cite{Kamilov_1998}. Therefore, up to frequencies of $\omega \sim 10^9$ 
s$^{-1}$, the product $\omega\tau\ll 1$, and the PM susceptibility 
can be considered purely real and frequency-independent.

Before discussing experimental verification, we note several practical considerations for the realization of the proposed FM/PM bilayer. The spectral modifications described in this work are driven by dipolar coupling between the FM and PM layers.  Interlayer exchange coupling, which is not accounted for in the present model, could suppress these effects. Experimentally, this can be mitigated by separating the FM and PM layers with a thin dielectric spacer. Furthermore, the model assumes an infinitely thick PM half-space. To ensure that the finite PM thickness does not artificially constrain the experimental results, its thickness should exceed the spin-wave penetration depth into the PM, which is on the order of the spin wavelength. Interface roughness can also affect the results via mechanisms similar to the ``orange-peel effect''~\cite{Schrag}. However, the magnitude of this effect is on the order of $h_r^2/(\lambda_r d)$, where $h_r$ and $\lambda_r$ are the amplitude and wavelength of the roughness profile, respectively. For $h_r^2/(\lambda_r d)\ll1$, this effect is negligible, which can be achieved with modern deposition techniques.

The results obtained in this work can be verified experimentally using Brillouin light spectroscopy (BLS) \cite{sebastian_micro-focused_2015}. As demonstrated in previous studies~\cite{Di_2015,Di_2015_2}, BLS can directly extract spin-wave spectra with high wave-vector resolution. By measuring these spectra as a function of the propagation direction, one can experimentally verify the predicted unidirectional propagation regime, frequency nonreciprocity, and the existence of forbidden angles. The findings of this work may also prove useful for creating controllable magnonic logic devices, such as a spin-wave diode.
\begin{acknowledgments}
The work was supported by State Contract No. FFUF-2024-0021 and the Foundation for the Advancement of Theoretical Physics and Mathematics ``BASIS'' (\#25-1-4-23-1).
\end{acknowledgments}

\bibliography{references}% Produces the bibliography via BibTeX.

\end{document}